\documentclass[fleqn,twoside]{article}
\usepackage{espcrc2}

\newcommand{ \be}{\begin{equation}}
\newcommand{ \ee}{\end{equation}}
\newcommand{ \bea}{\begin{eqnarray}}
\newcommand{ \eea}{\end{eqnarray}}
\newcommand{ \mysmall}[1]{\scriptscriptstyle #1} 

\newcommand{ \amu}{a_{\mu}}
\newcommand{ \mw}{M_{\mysmall{W}}}
\newcommand{ \mz}{M_{\mysmall{Z}}}
\newcommand{ \mh}{M_{\mysmall{H}}}
\newcommand{ \gev}  {\mbox{ GeV}}

\title{Status of the Standard Model Prediction of the Muon g-2}

\author{M. Passera\address{Dipartimento di Fisica, Universit\`a di Padova
        and INFN, Via Marzolo 8, 35131 Padova, Italy}\thanks{This work was
        supported in part by the European Program MRTN-CT-2004-503369.}}

\begin{document}

\begin{abstract}
The current status of the Standard Model prediction for the anomalous
magnetic moment of the muon is briefly reviewed and compared with the
present experimental value.
\end{abstract}

\maketitle
\setcounter{footnote}{0}

\section{Introduction}

Schwinger's 1948 calculation~\cite{Sch48} of the leading {\small QED}
contribution to the anomalous magnetic moment of the muon $\amu =
(g_{\mu}-2)/2$, equal to the one of the electron, was one of the very first
results of this theory, and one of its early confirmations. During the last
few years, in a sequence of increasingly precise measurements, the E821
Collaboration at Brookhaven has determined $\amu$ with a fabulous relative
precision of 0.5 parts per million (ppm)~\cite{BNL,BNL04,Lee}, serving as an
invaluable tool to test all sectors of the Standard Model ({\small SM}) and
to scrutinize viable alternatives to this theory~\cite{CM01}.  This note
provides a brief summary of the present status of the three contributions
into which the {\small SM} prediction $\amu^{\mysmall SM}$ is usually split
-- {\small QED}, electroweak and hadronic -- and a comparison with the
current experimental value.

\section{QED and Electroweak Contributions}

The {\small QED} contribution to $\amu$ arises from the subset of {\small
SM} diagrams containing only leptons ($e,\mu,\tau$) and photons. The
lowest-order contribution is $\amu^{\mysmall QED} (\mbox{1 loop}) =
\alpha/(2\pi)$~\cite{Sch48}.  Also the two- and three-loop {\small QED}
terms are known analytically -- see~\cite{MP04} for an update and a review
of these contributions. The four-loop term has been evaluated numerically, a
formidable task first accomplished by Kinoshita and his collaborators in the
early 1980s~\cite{KL81-KNO84}.  The latest analysis appeared
in~\cite{KN04}. Note that this four-loop contribution is about six times
larger than the present experimental uncertainty of $\amu$!  The evaluation
of the five-loop {\small QED} contribution is in progress~\cite{5loops}.

Adding up these terms, using the latest {\small CODATA}~\cite{CODATA}
recommended value for the fine-structure constant $\alpha^{-1} = 137.035
\,999 \,11 \,(46)$, known to 3.3 ppb, one obtains~\cite{MP04}
$
    \amu^{\mysmall QED} = 
    116 \, 584 \, 718.8 \, (0.3)\,(0.4)  \times 10^{-11}.       
$
The first error is mainly due to the uncertainty of the $O(\alpha^5)$
term, while the second one is caused by the 3.3 ppb uncertainty of the
fine-structure constant.

The electroweak ({\small EW}) contribution to $\amu$ is suppressed by a
factor $(m_{\mu}/\mw)^2$ with respect to the {\small QED} effects. The
one-loop part was computed in 1972 by several authors~\cite{ew1loop}:
$
     \amu^{\mysmall EW} (\mbox{1 loop}) = 
     \frac{5 G_{\mu} m^2_{\mu}}{24 \sqrt{2} \pi^2}
     \left[ 1+ \frac{1}{5}\left(1-4\sin^2\!\theta_{\mysmall{W}}\right)^2 
       + O(m^2_{\mu}/M^2_{\mysmall{Z,W,H}}) \right],
$
where $G_{\mu}=1.16637(1) \times 10^{-5}\gev^{-2}$. Employing the on-shell
definition $\sin^2\!\theta_{\mysmall{W}} =
1-M^2_{\mysmall{W}}/M^2_{\mysmall{Z}}$~\cite{Si80}, where
$\mz=91.1875(21)\gev$ and $\mw$ is the {\small SM} prediction of the $W$
mass (which can be derived, for example, from the simple formulae of
\cite{FOPS} leading to $\mw =80.383\gev$ for the Higgs mass $\mh=150\gev$),
we obtain $\amu^{\mysmall EW} (\mbox{1 loop}) = 194.8 \times 10^{-11}$.

The two-loop {\small EW} contribution to $\amu$ is not negligible because of
large factors of $\ln(M_{\mysmall{Z,W}}/m_f)$, where $m_f$ is a fermion mass
scale much smaller than $\mw$~\cite{KKSS}. It was computed in
1995~\cite{CKM95}.  The proper treatment of the contribution
of the light quarks was addressed in~\cite{PPD95KPPD02,CMV03}. These
refinements significantly improved the reliability of the fermionic part
(that containing closed fermion loops) of $\amu^{\mysmall EW}(\mbox{two
loop})$ leading, for $\mh=150\gev$, to
$
    \amu^{\mysmall EW} = 154(1)(2)\times 10^{-11}
$~\cite{CMV03}.
The first error is due to hadronic loop uncertainties, while the second one
corresponds to an allowed range of $\mh \in [114,250]\gev$, to the current
top mass uncertainty, and to unknown three-loop effects.  The leading-logarithm
three-loop contribution to $\amu^{\mysmall EW}$ is extremely
small~\cite{CMV03,DGi98}.
The results of \cite{HSW04} for the two-loop bosonic part of $\amu^{\mysmall
EW}$, performed without the large $\mh$ approximation previously employed,
agree with the previous evaluation \cite{CKM95} in the large Higgs mass limit.
Work is also in progress for an independent
recalculation based on the numerical methods of \cite{Topside}.

\section{The Hadronic Contribution}

The evaluation of the hadronic leading-order contribution
$\amu^{\mbox{$\scriptscriptstyle{HLO}$}}$, due to the hadronic vacuum
polarization correction to the one-loop diagram, involves long-distance
{\small QCD} for which perturbation theory cannot be employed. However,
using analyticity and unitarity, it was shown long ago that this term can be
computed from hadronic $e^+ e^-$ annihilation data via the dispersion
integral
$
      \amu^{\mbox{$\scriptscriptstyle{HLO}$}}= 
      \frac{1}{4\pi^3} \!
      \int^{\infty}_{4m_{\pi}^2} ds \, K(s) \sigma^{(0)}\!(s)
$~\cite{DISP},
where $\sigma^{(0)}\!(s)$ is the total cross section for $e^+ e^-$
annihilation into any hadronic state, with extraneous {\small QED}
corrections subtracted off. The kernel function $K(s)$ decreases
monotonically for increasing~$s$.

A prominent role among all $e^+ e^-$ annihilation measurements is played by
the precise data collected in 1994-95 by the {\small CMD-2} detector at the
{\small VEPP-2M} collider in Novosibirsk for the $e^+e^-\rightarrow
\pi^+\pi^-$ cross section at values of $\sqrt{s}$ between 0.61 and 0.96
${\rm GeV}$~\cite{CMD2-95}. The quoted systematic error of these data is
0.6\%, dominated by the uncertainties in the radiative corrections (0.4\%).
Recently~\cite{CMD2-98} the {\small CMD-2} Collaboration released its
1996-98 measurements for the same cross section in the full energy range
$\sqrt{s} \in [0.37,1.39]$ GeV. The part of these data for $\sqrt{s} \in
[0.61,0.96]$ GeV (quoted systematic error 0.8\%) agrees with their earlier
result published in~\cite{CMD2-95}. Also the {\small SND} Collaboration (at
the {\small VEPP-2M} collider as well) recently presented its analysis of
the $e^+e^-\rightarrow \pi^+\pi^-$ process for $\sqrt{s}$ between 0.39 and
0.98 ${\rm GeV}$, with a systematic uncertainty of 1.3\% (3.2\%) for
$\sqrt{s}$ larger (smaller) than 0.42 GeV~\cite{SND}. A hint of discrepancy,
at the level of the combined systematic error, occurs between the {\small
CMD-2} and {\small SND} measurements (the contribution to
$\amu^{\mbox{$\scriptscriptstyle{HLO}$}}$ of the {\small SND} data is a bit
higher than the corresponding one from {\small CMD-2})~\cite{CMD2-98}.
Further significant progress is expected from the new $e^+ e^-$ collider
{\small VEPP-2000} under construction in Novosibirsk~\cite{CMD2-98}.

In 2004 the {\small KLOE} experiment at the {\small DA$\Phi$NE} collider in
Frascati presented a precise measurement of $\sigma(e^+e^-\rightarrow
\pi^+\pi^-)$ via the initial-state radiation ({\small ISR}) method at the
$\phi$ resonance~\cite{KLOE-04}.  This cross section was extracted for
$\sqrt{s}$ between 0.59 and 0.97 GeV with a systematic error of 1.3\% and a
negligible statistical one.  There are some discrepancies between the
{\small KLOE} and {\small CMD-2} results ({\small KLOE}'s data lying higher
than the {\small CMD-2} fit below the $\rho$ peak, and lower on the peak and
above it), although their integrated contributions to
$\amu^{\mbox{$\scriptscriptstyle{HLO}$}}$ are similar~\cite{CMD2-98}. The
data of {\small KLOE} and {\small SND} disagree above the $\rho$ peak, where
the latter are significantly higher. The study of the $e^+e^-\rightarrow
\pi^+\pi^-$ process via the {\small ISR} method is also in progress at
{\small BABAR}~\cite{BABAR} and {\small BELLE}~\cite{BELLE}.  On the
theoretical side, analyticity, unitarity and chiral symmetry provide strong
constraints for the pion form factor in the low-energy region~\cite{Chiral}.

The evaluations of the dispersive integral based on the
analysis~\cite{CMD2-95} of the 1994-95 {\small CMD-2} data are in good
agreement:\footnote{The evaluation of~\cite{ELZ04} is not included as its
result is being revised~\cite{ZeninPrivate}.}
\bea
\label{eq:DEHZ04}
      \mbox{\cite{DEHZ04}} &
      \amu^{\mbox{$\scriptscriptstyle{HLO}$}} \,\,= & 
      6934 \, (53)_{exp} (35)_{rad} \times 10^{-11},  \nonumber \\[-1mm]
\label{eq:J03}
      \mbox{\cite{J03}}    &
      \amu^{\mbox{$\scriptscriptstyle{HLO}$}} \,\,= & 
      6948 \, (86) \times 10^{-11},  \nonumber \\[-1mm]
\label{eq:HMNT03}
      \mbox{\cite{HMNT03}} &
      \amu^{\mbox{$\scriptscriptstyle{HLO}$}} \,\,= & 
      6924 \, (59)_{exp} (24)_{rad} \times 10^{-11},  \nonumber \\[-1mm]
\label{eq:dTY04}
      \mbox{\cite{dTY04}}  &
      \amu^{\mbox{$\scriptscriptstyle{HLO}$}} \,\,= & 
      6944 \, (48)_{exp} (10)_{rad} \times 10^{-11}. \nonumber 
\eea
Reference \cite{DEHZ04} updates \cite{DEHZ03} and already includes {\small
KLOE}'s results. The recently released data of {\small CMD-2}~\cite{CMD2-98}
and {\small SND}~\cite{SND} are not yet included.

The authors of \cite{ADH98} pioneered the idea of using vector spectral
functions derived from the study of hadronic $\tau$ decays (see
\cite{TauRevs,DHZ05} for recent reviews) to improve the evaluation of the
dispersive integral.  Indeed, assuming isospin invariance to hold, the
isovector part of the cross section for $e^+e^-\rightarrow$ hadrons can be
calculated via the Conserved Vector Current relations from $\tau$-decay
spectra.  The latest analysis with {\small ALEPH}~\cite{ALEPH}, {\small
CLEO}~\cite{CLEO}, and {\small OPAL}~\cite{OPAL} data yields
$\amu^{\mbox{$\scriptscriptstyle{HLO}$}} = 7110 \,
(50)_{exp} (8)_{rad} (28)_{SU(2)} \times 10^{-11}
$\cite{DEHZ03}.
Isospin-breaking corrections were applied~\cite{MS88CEN}. Information from
$\tau$ decays was also included in one of the analyses of~\cite{dTY04},
leading to
$
      \amu^{\mbox{$\scriptscriptstyle{HLO}$}} =
      7027 \, (47)_{exp} (10)_{rad} \times 10^{-11}.
$

Although the precise {\small CMD-2} $e^+e^-\rightarrow \pi^+\pi^-$
data~\cite{CMD2-95} are consistent with the corresponding $\tau$ ones for
energies below $\sim0.85$ GeV, they are significantly lower for larger
energies. {\small KLOE}'s $\pi^+\pi^-$ spectral function confirms this
discrepancy with the $\tau$ data; on the contrary, the recent {\small SND}
results are compatible with them~\cite{SND}.  This discrepancy could be
caused by inconsistencies in the $e^+e^-$ or $\tau$ data, or in the
isospin-breaking corrections which must be applied to the latter. Indeed,
the mentioned disagreements between the $e^+e^-$ data sets need careful
consideration. On the other hand, in spite of the agreement of the $\tau$
data sets~\cite{DHZ05}, the question remains whether all possible
isospin-breaking effects have been properly taken into
account~\cite{DEHZ04,eetau}.

The hadronic higher-order contribution can be divided into two parts:
$
     \amu^{\mbox{$\scriptscriptstyle{HHO}$}}=
     \amu^{\mbox{$\scriptscriptstyle{HHO}$}}(\mbox{vp})+
     \amu^{\mbox{$\scriptscriptstyle{HHO}$}}(\mbox{lbl}).
$
The first term is the $O(\alpha^3)$ contribution of diagrams containing
hadronic vacuum polarization insertions~\cite{Kr96}. Its latest value is
$\amu^{\mbox{$\scriptscriptstyle{HHO}$}}(\mbox{vp})= -97.9 \, (0.9)_{exp}
(0.3)_{rad} \times 10^{-11} $~\cite{HMNT03}, obtained using $e^+ e^-$
annihilation data; it changes by $\sim -3\times 10^{-11}$ if hadronic
$\tau$-decay data are used instead~\cite{DM04}.
The second term, also of $O(\alpha^3)$, is the hadronic light-by-light
contribution; as it cannot be determined from data, its evaluation relies on
specific models.  In 2001 the authors of~\cite{KN01KNPdR01} uncovered a sign
error in earlier evaluations of its dominating pion-pole part. Their
estimate, based also on previous results for the quark and charged-pions
loop parts~\cite{HK01BPP01}, is
$
      \amu^{\mbox{$\scriptscriptstyle{HHO}$}}(\mbox{lbl}) =  
      80\,(40)\times 10^{-11}
$~\cite{Ny03}.
A higher value was obtained in 2003 including short-distance
{\small QCD} constraints:
$
      \amu^{\mbox{$\scriptscriptstyle{HHO}$}}(\mbox{lbl}) =
      136\,(25)\times 10^{-11}
$~\cite{MV03}.
Further independent calculations would provide an important check of this
result for $\amu^{\mbox{$\scriptscriptstyle{HHO}$}}(\mbox{lbl})$, a
contribution whose uncertainty may become the ultimate limitation of the
{\small SM} prediction of the muon $g$$-$$2$.

\section{Standard Model vs.\ Measurement}

The first column of Table 1 shows
$
    \amu^{\mysmall SM} = 
         \amu^{\mysmall QED} \!+\!
         \amu^{\mysmall EW}  \!+\!
	 \amu^{\mbox{$\scriptscriptstyle{HLO}$}} \!+\!
	 \amu^{\mbox{$\scriptscriptstyle{HHO}$}}.
$
The values employed for $\amu^{\mbox{$\scriptscriptstyle{HLO}$}}$ are
indicated by the reference on the left (\cite{dTY04} quotes two values, see
Sec.\ 3); all $\amu^{\mysmall SM}$ values were derived with
$\amu^{\mbox{$\scriptscriptstyle{HHO}$}}(\mbox{lbl})\!=\! 136\,(25)\times
10^{-11}$~\cite{MV03}. Errors were added in quadrature.
\begin{table}
{\small
\begin{tabular}{lll}
\hline 
$\amu^{\mbox{$\scriptscriptstyle{SM}$}} \times 10^{11}$ & 
$\Delta \times 10^{11}$  & ~~$\sigma$           \\
\hline 
\mbox{\cite{DEHZ04}}~~116\,591\,845 (69)~& 
                      235 (91) & ~~2.6 $\langle3.0\rangle$  \\
\mbox{\cite{J03}}~~116\,591\,859    (90)~& 
                      221 (108)& ~~2.1 $\langle2.5\rangle$  \\
\mbox{\cite{HMNT03}}~~116\,591\,835 (69)~& 
                      245 (91) & ~~2.7 $\langle3.1\rangle$  \\
\mbox{\cite{dTY04}}~~116\,591\,855  (55)~& 
                      225 (81) & ~~2.8 $\langle3.2\rangle$  \\
\mbox{\cite{DEHZ03}}~~116\,592\,018 (63)~&
                      \phantom{0}62 (87)& ~~0.7 $\langle1.3\rangle$\\
\mbox{\cite{dTY04}}~~116\,591\,938  (54)~ & 
                      142 (81) & ~~1.8 $\langle2.3\rangle$ \\
\hline
\end{tabular}\\[1mm]
}
Table 1: Standard Model vs.\ measurement.
\vspace{-6mm}
\end{table} 
%
%
%
%
%
The present world average experimental value for the muon $g$$-$$2$
is
$
    \amu^{\mbox{$\scriptscriptstyle{EXP}$}}  = 
               116 \, 592 \, 080 \, (60) \times 10^{-11} 
	       ~(0.5~\mbox{ppm})
$~\cite{BNL04}.
The differences $\Delta =\amu^{\mbox{$\scriptscriptstyle{EXP}$}}-
\amu^{\mbox{$\scriptscriptstyle{SM}$}}$ are listed in the second column of
Table 1, while the numbers of ``standard deviations'' ($\sigma$) appear in
the third one. Higher discrepancies, shown in angle brackets, are obtained
if $\amu^{\mbox{$\scriptscriptstyle{HHO}$}}(\mbox{lbl}) = 80\,(40)\times
10^{-11}$~\cite{Ny03} is used instead of $136\,(25)\times
10^{-11}$~\cite{MV03}.

\section{Conclusions}

The discrepancies between recent {\small SM} predictions of $\amu$ and the
current experimental value vary in a very wide range, from 0.7 to 3.2
$\sigma$, according to the values chosen for the hadronic contributions.  In
particular, the leading-order hadronic contribution depends on which of the
two sets of data, $e^+e^-$ collisions or $\tau$ decays, are employed.
The puzzling discrepancy between the $\pi^+\pi^-$ spectral functions from
$e^+e^-$ and isospin-breaking-corrected $\tau$ data could be caused by
inconsistencies in the $e^+e^-$ or $\tau$ data, or in the isospin-breaking
corrections applied to the latter. Indeed, disagreements occur between
$e^+e^-$ data sets, requiring further detailed investigations. On the other
hand, $\tau$ data sets are in agreement, but their connection with the
leading hadronic contribution to $\amu$ is less direct, and one wonders
whether all possible isospin-breaking effects have been properly taken into
account.  Using $e^+e^-$ data, the {\small SM} prediction of the muon
$g$$-$$2$ deviates from the present experimental value by 2--3~$\sigma$.

The impressive results of the {\small E821} experiment are still limited by
statistical errors.  A new experiment, {\small E969}, has been approved (but
not yet funded) at Brookhaven in 2004~\cite{Lee,E969}. Its goal is to reduce
the present experimental uncertainty by a factor of 2.5 to about 0.2 ppm. A
letter of intent for an even more precise $g$$-$$2$ experiment was submitted
to {\small J-PARC} with the proposal to reach a precision below 0.1
ppm~\cite{JPARC}.  While the {\small QED} and {\small EW} contributions
appear to be ready to rival these precisions, much effort will be needed to
reduce the hadronic uncertainty by a factor of two. This effort is
challenging but possible, and certainly well motivated by the excellent
opportunity the muon $g$$-$$2$ is providing us to unveil (or constrain)
``new physics'' effects.
%

{}~

\noindent {\bf Acknowledgments.} I would like to thank the organizers for
their kind invitation and excellent organization of this workshop, and
S.~Eidelman and B.L.~Roberts for very useful communications.


\end{document}